\documentclass[letter]{aa}

\newcommand{\argmax}{\mathop{\mathrm{argmax}}\limits}
\usepackage{graphicx}
\usepackage{txfonts}
\begin{document}

   \title{CS Cha B: A disc-obscured M-type star mimicking a polarised planetary companion \thanks{The extracted spectrum of CS Cha B is available in electronic form
at the CDS via anonymous ftp to cdsarc.u-strasbg.fr (130.79.128.5)
or via http://cdsweb.u-strasbg.fr/cgi-bin/qcat?J/A+A/}}

   \author{S. Y. Haffert \thanks{NASA Hubble fellow}
          \inst{1}
          \and
          R. G. van Holstein\inst{2,3}
          \and
          C. Ginski\inst{4}
          \and
          J. Brinchmann\inst{5}
          \and
          I. A. G. Snellen\inst{2}
          \and
          J. Milli\inst{6}
          \and
          T. Stolker\inst{7}
          \and
          C. U. Keller\inst{2}
          \and
          J. Girard\inst{8}
          }

      \institute{
            Steward Observatory, Unversity of Arizona, 933 North Cherry Avenue, Tucson, Arizona\\
            \email{shaffert@arizona.edu}
         \and
             Leiden Observatory, Leiden University, PO Box 9513, Niels Bohrweg 2, 2300 RA Leiden, The Netherlands
        \and
            European Southern Observatory, Alonso de C\'{o}rdova 3107, Casilla 19001, Vitacura, Santiago, Chile 
        \and
            Astronomical Institute Anton Pannekoek, University of Amsterdam, PO Box 94249, 1090 GE Amsterdam, The Netherlands
        \and
            Instituto de Astrofísica e Ciências do Espaço, Universidade do Porto, CAUP, Rua das Estrelas, PT4150-762 Porto, Portugal
        \and
            Universit\'e Grenoble Alpes, CNRS, IPAG, 38000 Grenoble, France 
        \and
            Institute for Particle Physics and Astrophysics, ETH Zurich, Wolfgang-Pauli-Strasse 27, 8093 Zurich, Switzerland
         \and 
            Space Telescope Science Institute, Baltimore 21218, MD, USA
         }
 
   \date{Received 19 June 2020; accepted 7 July 2020}

 
  \abstract
   {Direct imaging provides a steady flow of newly discovered giant planets and brown dwarf companions. These multi-object systems can provide information about the formation of low-mass companions in wide orbits and/or help us to speculate about possible migration scenarios. Accurate classification of companions is crucial for testing formation pathways.
   }
   {In this work we further characterise  the recently discovered candidate for a planetary-mass companion CS Cha b and determine if it is still accreting.
   }
   {MUSE is a four-laser-adaptive-optics-assisted medium-resolution integral-field spectrograph in the optical part of the spectrum. We observed the CS Cha system to obtain the first spectrum of CS Cha b. The companion is characterised by modelling both the spectrum from 6300 \AA\, to 9300 \AA\, and the photometry using archival data from the visible to the near-infrared (NIR).
   }
   {We find evidence of accretion and outflow signatures in H$\alpha$ and OI emission. The atmospheric models with the highest likelihood indicate an effective temperature of $3450\pm50$ K with a $\log{\mathrm{g}}$ of $3.6 \pm 0.5$ dex. Based on evolutionary models, we find that the majority of the object is obscured. We determine the mass of the faint companion with several methods to be between 0.07 $M_\odot$ and 0.71 $M_\odot$ with an accretion rate of $\dot{M}=4\times10^{-11\pm0.4} M_{\odot}\mathrm{yr}^{-1}$.
   }
   {Our results show that CS Cha B is most likely a mid-M-type star that is obscured by a highly inclined disc, which has led to its previous classification using broadband NIR photometry as a planetary-mass companion. This shows that it is important and necessary to observe over a broad spectral range to constrain the nature of faint companions.
    }

   \maketitle

\section{Introduction}
The direct-imaging instruments GPI and SPHERE \citep{macintosh2014gpi, beuzit2019sphere} have been used to discover and characterise several substellar companions \citep{bailey2014hd106906, keppler2018discovery, bohn2020tyc8998} on wide orbits ($\geq10$ AU). Many of these are thought to be brown dwarfs, and only a fraction of these companions are actual, confirmed exoplanets \citep{guenther2005gqlup, lafreniere2008rxsj1609,biller2010pztel, keppler2018discovery, haffert2019pds70}. One of the major puzzles of these wide-orbit substellar companions is the process that is responsible for their formation. There are three major theories for the formation of substellar companions in circumstellar discs. The first is a top-down approach where the circumstellar disc fragments due to gravitational instabilities (GI), and the fragments collapse into protoplanetary cores \citep{boss1997GI}. The second is the core accretion (CA) model, where small planetesimals cluster together and form protoplanetary cores which then scatter to large separations \citep{pollack1996coreaccretion}. In the third formation mechanism, the companion forms during the initial collapse of the prestellar core, where the clump of gas and dust breaks up into separate cores \citep{hennebelle2008core}. These three formation mechanisms produce different signatures in the companion properties. This difference has been used to estimate the contribution of the different formation mechanisms to the observed companion distribution \citep{wagner2019wideoccurence}.

Accurate determination of the parameters of these companions is fundamental to assess the efficiency of the different formation pathways. But deriving the fundamental parameters of these companions from spectral energy distributions (SEDs) based on a few photometric measurements can be degenerate among the parameters and may lead to large uncertainties in the estimates of their mass, temperature, and radius. Detailed characterisation of these substellar companions is therefore necessary to link planet formation to the observations. Here we focus on the characterisation of the young stellar system CS Cha and its wide-orbit companion.

CS Cha is a classical T-Tauri type object \citep{appenzeller1977yso, manara2014xshooter} and likely a spectroscopic binary ($\sim4$ AU separation) \citep{guenther2007cscha} with an estimated age of 2 $\pm$ 2 Myr \citep{luhman2008chameleon} located in the Chamaeleon I association at a distance of 176 $\pm$ 3 parsec \citep{gaia2018dr2}. The infrared (IR) SED of CS Cha is known to contain a large amount of IR excess, with a lack of emission at 10 $\mathrm{\mu}$m \citep{gauvin1992cha}. A large cavity in the circumbinary disc can explain this dip. Near-infrared (NIR) polarimetry with SPHERE/IRDIS has spatially resolved the disc but not the cavity, which is likely behind the coronagraphic mask \citep[hereafter G18]{ginski2018cscha}. A surprising finding was the discovery of a possible planetary companion, CS Cha b, at a separation of 228.8 AU. The companion discovered by G18 was found to be very strongly linearly polarised. A highly inclined circumplanetary disc was necessary to explain the SED and the high degree of linear polarisation. An upper limit of 20 M$_\mathrm{J}$ was determined for the mass of the companion based on the measured polarisation and SED.

In this letter we present the first spatially resolved spectroscopy of CS Cha b with the optical integral-field spectrograph Multi Unit Spectroscopic Explorer (MUSE). With the addition of the optical spectrum of CS Cha b we are able to refine its fundamental parameters, showing that it is very likely an M-type star that is highly obscured by its own circumstellar disc rather than a planetary-mass companion.

\section{Observations and spectrum extraction}
We used MUSE \citep{bacon2010muse} at the Very Large Telescope (VLT) to observe the CS Cha system on April 19 2019; MUSE is an integral-field unit that can be fed by the new Laser Tomographic Adaptive Optics (LTAO) \citep{oberti2016ltao,madec2018aof} system at UT4. The LTAO system delivers near diffraction-limited performance at optical wavelengths. The combination of the LTAO system with the 25 mas spatial pixel (spaxels) size in the Narrow Field Mode (NFM) enables MUSE to make high-resolution observations. The atmospheric conditions during the observations were excellent, the seeing at 500 nm was between 0.25" and 0.5", and most of the time 0.3" or better. The coherence time, $\tau_0$, was between 10 and 20 ms. These conditions provided a high-quality dataset.

The raw data were processed and calibrated with the MUSE pipeline in ESOREX version 2.8.2 \citep{weilbacher2014musedpr}. The pipeline produces absolutely calibrated photometric datacubes. The point spread function (PSF) of MUSE is not Nyquist sampled at 25mas. Therefore interpolation will lead to artifacts if used as a way to apply subpixel shifts. For that reason we decided to only apply full-pixel shifts. This will degrade the full width at half maximum (FWHM), but we will be less susceptible to interpolation errors. We started the stacking procedure by first creating a white-light image from the 3D cube. We then determined the subpixel centre for each exposure by calculating the centre of gravity. We took the pixel that was closest to the centre of gravity as the reference pixel for stacking.

\subsection{Post-processing to remove the stellar halo}
\begin{figure}
\centering
\includegraphics[width=\columnwidth]{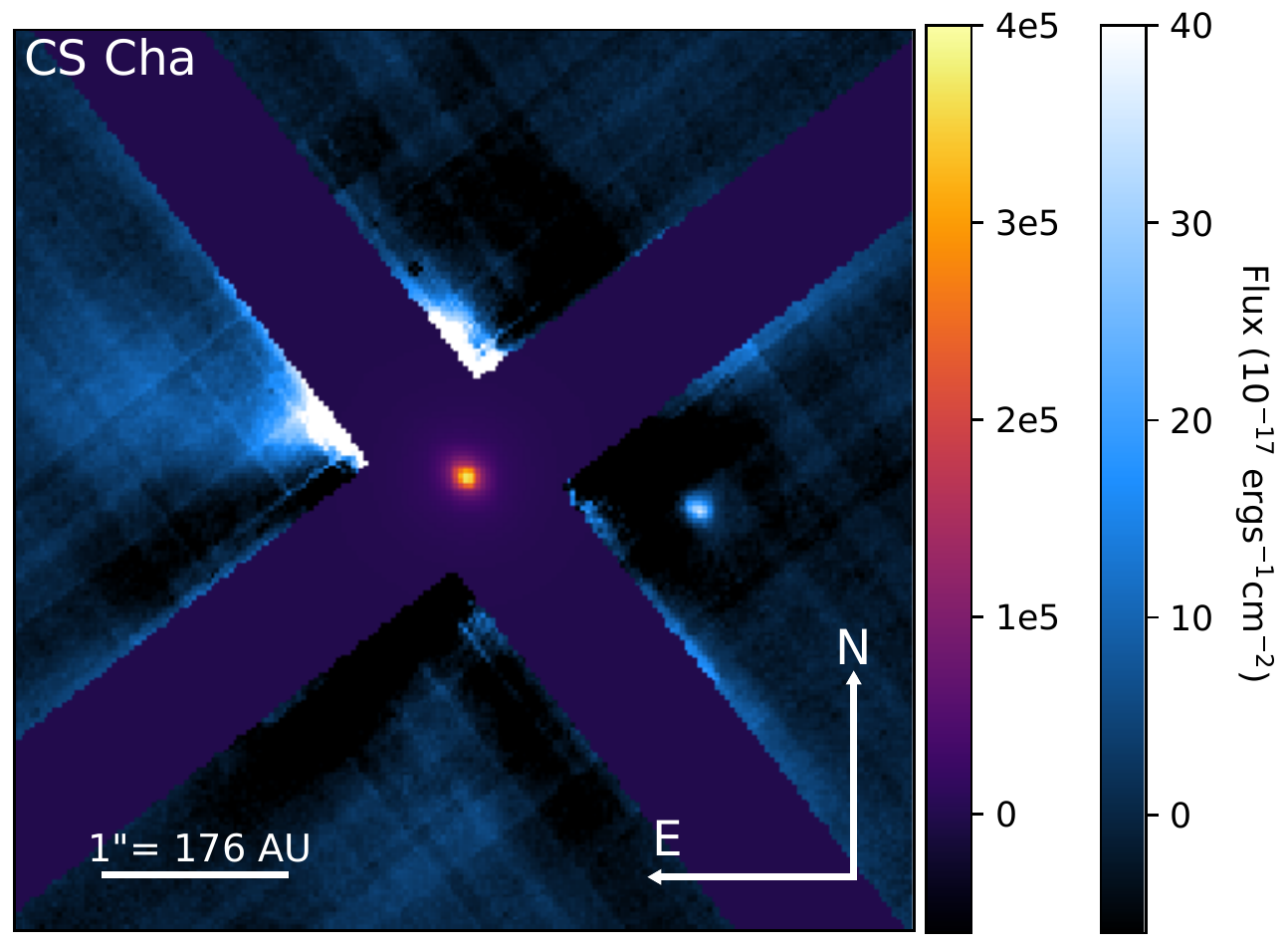}

\caption{Combined flux image integrated from 6400 \AA\,to 9300 \AA\, of CS Cha before (orange) and after (blue) removal of the halo of the primary component. The reddish colours show the primary component, while the blue-white colours show the post-processed image. Two crossed bars are used to mask the IFU slices that contain the primary and the diffraction structure of the spiders to enhance the dynamic range of the figure. The primary component is roughly ten magnitudes brighter than CS Cha b. CS Cha b is found at the expected position of 1.3 arcseconds west of the primary. }
\label{fig:system_image}
\end{figure}

To increase the sensitivity to faint companions we needed to subtract all stellar light from the primary. Previous methods to detect faint companions with MUSE were targeting single emission lines \citep[e.g.][]{haffert2019pds70}. The high-resolution spectral differential imaging technique is not suitable to retrieve the continuum of the companion because the continuum of the companion is removed in favour of better speckle subtraction at the position of the emission line. A different post-processing algorithm is necessary to retrieve the continuum. Therefore we removed the stellar halo by subtracting a radial profile. The radial profile was measured in radial steps of 12.5 mas. The companion is detected in the white light-image as shown in Fig. \ref{fig:system_image}.

We determined the position of the companion by measuring the local center of gravity in the white-light image after subtraction of the radial profile. The measured angular separation is 1.31" $\pm$ 0.025", and the measured PA is 261.3$^\circ\pm1^\circ$. The error margins are derived from the assumption of a one-pixel astrometric accuracy. Our measurements are in agreement with the astrometry of G18 and reaffirm that the object is co-moving with the primary component.

\subsection{Extraction of the companion spectrum}
To extract the flux of the companion we used a shifted and normalised version of the stellar PSF as a template for the companion PSF. The flux is estimated by taking a weighted sum within a 150 milli-arcsecond aperture with the template PSF as weights. This automatically accounts for flux loss outside the pixels that are fitted and applies an optimally weighted summation of the flux within the aperture pixels.

\begin{figure*}[ht]
\includegraphics[width=\textwidth]{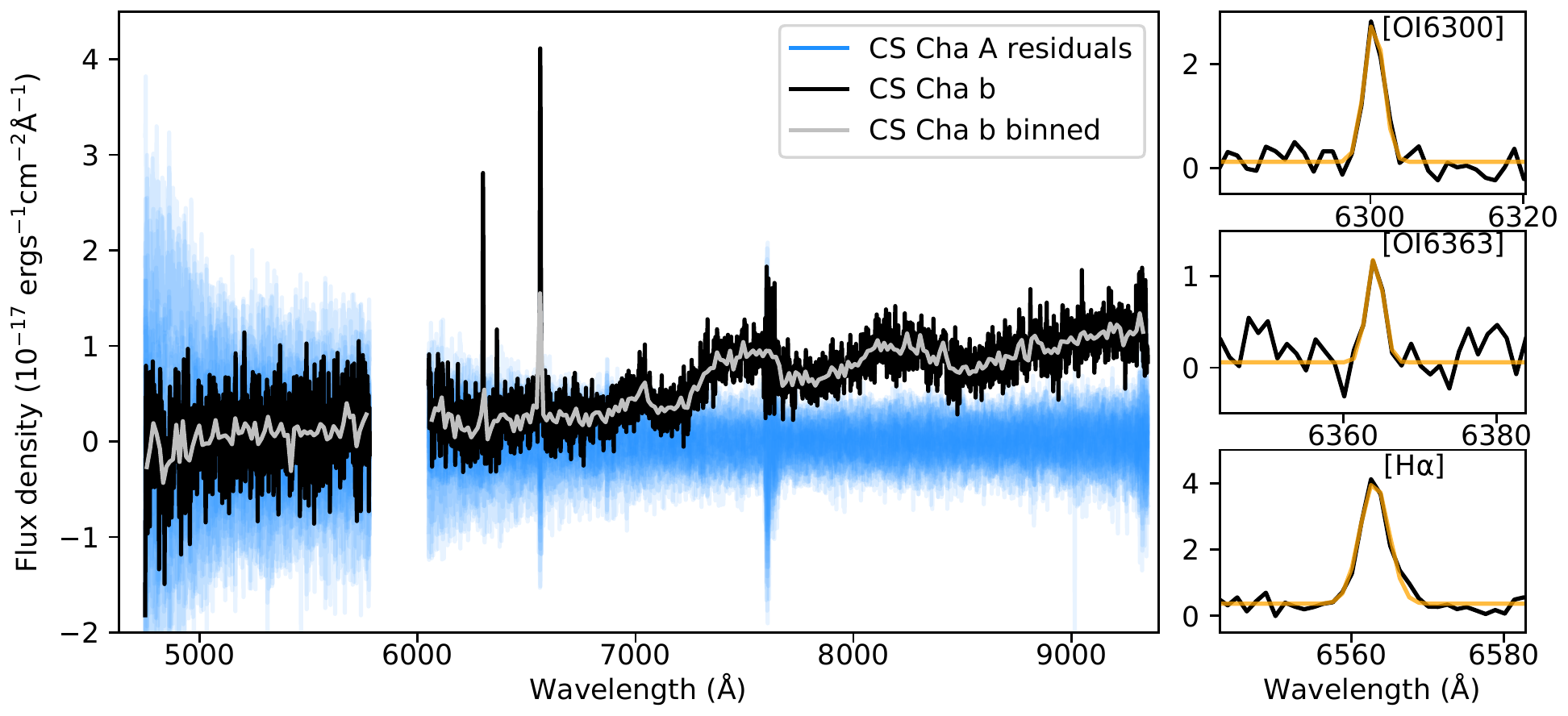}

\caption{Extracted spectrum of CS Cha b is shown in black, and the stellar residuals at several positions around CS Cha b are shown in blue. The grey line is a binned-down version of the spectrum to enhance S/N. Even after binning, no clear absorption features are present. The spectrum of CS Cha b shows weak H$\alpha$ emission and two forbidden OI lines at 6300 \AA\, and 6364 \AA, which are highlighted on the right. The orange lines show the Gaussian fit to each emission line. The OI 6300 has a line strength that is nearly equal to the H$\alpha$ emission line. At 7600 \AA\,there are spurious features due the presence of imperfect correction of strong telluric lines. The gap between 5800 \AA\, and 6051 \AA\, is due to the presence of a notch filter, which is centred on the sodium doublet in the LTAO system that blocks the sodium-based laser guide star. }

\label{fig:extracted_spectrum}

\end{figure*}

After the radial subtraction there are still some slowly varying residuals at the position of the companion, which can be seen in Fig. \ref{fig:system_image}. These residual features are removed by fitting a second-order 2D polynomial. The residual spectra at several positions around CS Cha b were used to estimate the values for the polynomial fit. The aperture at the position of CS Cha b was excluded from the background fit. This second step strongly suppresses the leftover stellar residuals. The final extracted spectrum and post-processed residuals can be seen in Fig.  \ref{fig:extracted_spectrum}.

The absolute flux density that we extract is consistent with the HST photometry \citep{ginski2018cscha}. The F814W filter photometry is slightly brighter than the MUSE spectrum. This is because CS Cha b is positioned directly on top of the spider diffraction structure in the HST images, which is also why it was not detected before \citep{ginski2018cscha}. The addition of the spider residuals increases the effective measured flux, which explains the discrepancy between the observations. The F606W point is an upper-limit because CS Cha was not detected in that filter. The MUSE observations likewise do not have sufficient signal-to-noise ratio (S/N) to measure the continuum below about 6100 \AA.

\begin{table}
\caption{Derived posterior parameters of the model parameters from the MCMC chain.}
\label{tab:emcee}
\centering
\begin{tabular}{c c c}
\hline
\hline
Model parameter & Median & comment \\ \hline
$T_{\mathrm{eff}}$ & $3452^{+44}_{-74}$ K &\\
$\log{g}$ & $3.66^{+0.43}_{-0.52}$ & \\
$R_{\mathrm{fit}}$  & $0.038 \pm 0.0024$ $R_\mathrm{\odot}$ & \\
$A_\mathrm{V}$ & $0.54 \pm0.3$ &  \\
$R_\mathrm{V}$ & $\ge$ 6 & 1 $\sigma$ lower limit \\
$v_r$ & $-62^{+59}_{-77}$ kms$^{-1}$ & \\
\hline
\end{tabular}
\end{table}

\section{Analysis and characterisation}
\subsection{Detection of accretion and outflows in CS Cha b}

\begin{table}
\caption{Emission line parameters.}
\label{tab:line}
\centering
\begin{tabular}{lccc}
\hline
\hline
Line & Flux & FW10 & EW \\
 & $($ ergs$^{-1}$cm$^{-2})$ &  $($kms$^{-1})$  & $($\AA$)$\\
\hline
OI 6300 & $8.5\pm1.6\cdot 10^{-17}$ & $254\pm38$ & $-71\pm66$ \\
OI 6364 & $3.1\pm1.5\cdot 10^{-17}$  & $217\pm79$ & $-53\pm96$ \\
H$\alpha$ & $17.3\pm2.1\cdot 10^{-17}$ & $369\pm35$ & $-47\pm17$ \\
\hline
\end{tabular}
\tablefoot{The full width at 10\%\ of the maximum (FW10) has not been corrected for instrumental broadening, which are 224 kms$^{-1}$ at 6300 \AA\, and 211 kms$^{-1}$ at 6562.8 \AA\, \citep{bacon2017wfmmuse}. The equivalent width (EW) is negative for emission lines. }
\end{table}

Surprisingly, we detect the presence of strong [OI] lines at 6300.8 \AA\,and 6363.8 \AA\, relative to the H$\alpha$ line in the spectrum of CS Cha b. These emission lines spatially coincide with the position of the companion, and therefore this emission probably originates from the companion itself or its immediate environment. The forbidden oxygen lines are usually assumed to be caused by the presence of low-gas-density outflows \citep{appenzeller1984forbiddenwinds, edwards1987diskwind}. The line flux is estimated by fitting a Gaussian to each of the three lines. We use the full width at 10\%\ of the maximum (FW10), local continuum, and line flux as free parameters in the Gaussian model. The measured properties of each fit can be found in Table \ref{tab:line}. From the line fluxes we derive a line ratio close to 1/2 for [OI6300] / H$\alpha$ and 1/6 for [OI6364] / H$\alpha$. The [OI6300] / [OI6364] is close to 3, which is expected based on the theoretical line ratio \citep{storey2000oiratio}.

From the FW10 of the H$\alpha$ line we conclude that the emission originates from accretion. Empirically the threshold has been a FW10 of at least 200 or 270 kms$^{-1}$ to decide whether the object exhibits accretion or photospheric activity \citep{jayawardhana2003accretion, white2003accretion, natta2004accretionwidth}. The measured FW10 still includes instrumental line broadening. We estimate the intrinsic FW10 by subtracting the instrumental FW10 \citep{bacon2017wfmmuse} in quadrature, $\mathrm{FW}10 = \sqrt{\mathrm{FW}10_{obs}^2 - \mathrm{FW}10_{ins}^2}$. After these corrections the intrinsic FW10 of CS Cha b is $298\pm43$ kms$^{-1}$, which is higher than either of the cutoff criteria for accretion. We therefore conclude that CS Cha b is actively accreting.

For comparison, FW Tau b, which is also an accreting companion in a wide orbit with an ouflow, has an [OI6300] / H$\alpha$ line ratio close to 1/4 and a [OI6364] / H$\alpha$ line ratio of almost 1/10 \citep{bowler2014fwtau}. The line ratios of CS Cha b are quite similar to those of FW Tau b. This suggests that CS Cha b is most likely an accreting companion with an outflow.

\subsection{Broadband properties of CS Cha b}
While the detected broadband integrated flux is quite faint ($\sim21$st magnitude), the optical spectrum does not resemble that of a young late-type object \citep{henry1994mdwarfs,kirkpatrick1995mdwarfs,kirkpatrick1999Ltype}, as suggested by G18. This indicates that the object is likely to have a higher temperature than previously thought. We compare the measured spectrum and photometry to the BTSettl model spectra \citep{baraffe2015btsettl} to estimate the effective temperature, surface gravity, and mass of the companion. The reddening model from \citet{cardelli1989extinction} is used to account for interstellar extinction. This model is a two-parameter extinction model, $A=A_v\left(f(\lambda) + g(\lambda)/R_v\right)$, where $A_v$ is the strength of the extinction, $R_v$ the reddening, and $f$ and $g$ are empirically calibrated polynomials. 

The Markov Chain Monte Carlo sampler emcee \citep{foreman2013emcee} was used to derive the posterior distribution of the model parameters. The definition of our log-likelihood function can be found in Appendix \ref{ap:mcmc}. We used uniform priors for the effective temperature $T_{\mathrm{eff}}$ [2000 K to 4000 K], surface gravity $\log{g}$ [2.5 dex to 5.5 dex], radius $R_{\mathrm{fit}}$ [0.01 $R_\mathrm{\odot}$ to 0.35 $R_\mathrm{\odot}$], extinction $A_\mathrm{V}$ [0.01, 10] and reddening $R_\mathrm{V}$ [0.1, 25] and the radial velocity $v_r$ [-500 kms$^{-1}$, 500 kms$^{-1}$]. The prior of the distance to the system was taken as a Gaussian distribution with mean 176 pc and standard deviation 0.5 pc \citep{gaia2018dr2}. The radius and distance are used to scale the model flux density by $(R/d)^2$. We can accurately derive the flux scaling by fitting $R$ because the distance is well determined by Gaia. The BTSettl model has been evaluated on a model grid with steps of 0.5 dex for the surface gravity and 100 $K$ for the effective temperature. We use bilinear interpolation to create subgrid spectra and down-sampled the model spectra to a resolving power of 3000. The first 1500 of the 10 000 samples of each MCMC chain were discarded to account for the burn-in effects during the convergence. The resulting model parameters are summarised in Table \ref{tab:emcee}.

\begin{figure*}[ht]
1\includegraphics[width=\textwidth]{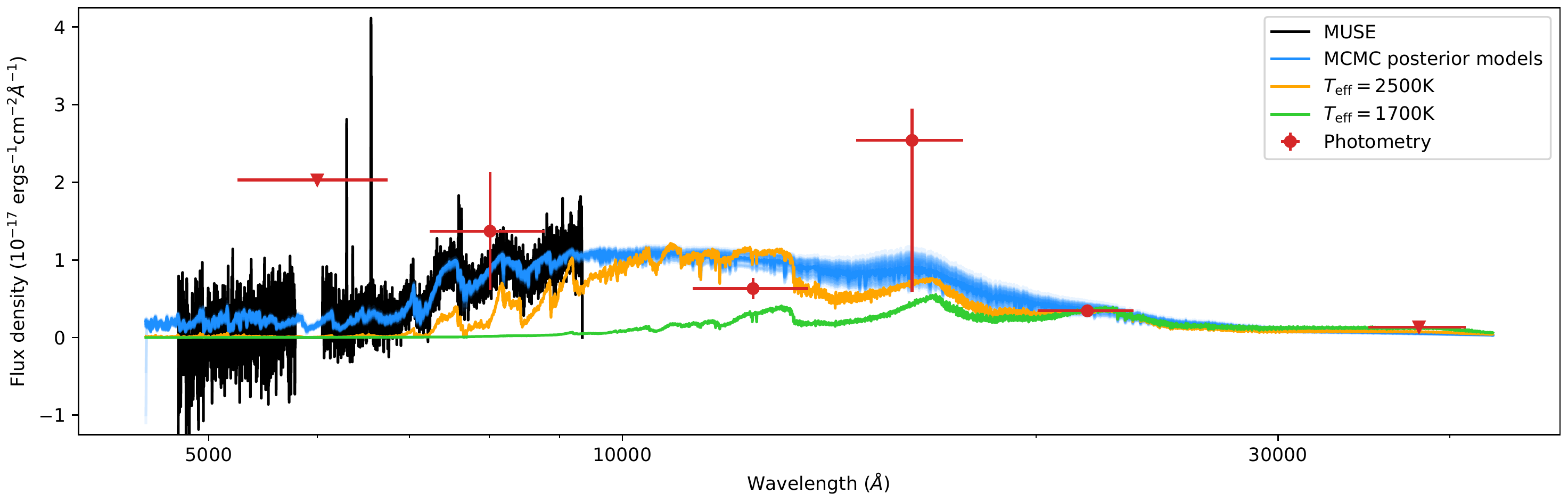}
\caption{Observations of CS Cha b and the best fit MCMC model. The optical spectrum from MUSE is black while the photometric observations (listed in Table \ref{tab:obs}.) are shown in red. The vertical bars of the photometric points indicate the uncertainty in the flux and the horizontal bars show the width of the filter that was used. The triangle markers are used for upper limits. The blue lines show 100 random samples from the derived posterior distributions for which median temperature is T$_{\mathrm{eff}}$=3452 K. A T$_{\mathrm{eff}}$=2500 K (orange) and T$_{\mathrm{eff}}$=1700 K (green) model are shown for comparison. The lower temperature models have been scaled to match the K-band photometry. The low temperature models have a significantly lower flux in the optical part compared to the observations. }
\label{fig:emcee_fit_spectrum}
\end{figure*}

Figure \ref{fig:emcee_fit_spectrum} shows the extracted spectrum and the photometric data points overlayed with 100 random spectra from the MCMC sampling chains. The model derived from the MCMC chains fit both the optical spectrum and the photometry. The observed spectrum is also compared with 2500 K and 1700 K models, which were the previously proposed temperatures \citep{ginski2018cscha}. Both low-temperature models are shown without extinction and were scaled to match the K-band photometry. The data show that there is a strong preference for models with higher temperature because there is a significant amount of optical flux, which cannot be explained by the lower temperature models.

\subsection{Determining the nature of CS Cha b}
The most striking result is the effective temperature of $3452^{+44}_{-74}$ K for CS Cha b, which is much higher than what has previously been proposed based on the NIR photometry. A second remarkable point is the small radius  $R_{\mathrm{fit}}=0.038\pm0.0024 R_\mathrm{\odot}$ that is preferred by the MCMC chain. This radius is unphysical but is necessary to fit the faintness of this object. The radius for a 3452 K object with an age of 2 Myr according to the evolutionary track of the BTSettle models is $R_{\mathrm{evo}}=1.26\pm0.04$ $R_{\odot}$, which is more than an order of magnitude larger.

G18 suggested that a highly inclined disc could explain the high degree of linear polarisation at NIR wavelengths. Such a disc geometry could also obscure the object itself. The majority of the polarised signal comes from forward and backward scattering from the surface of the disc. If the disc blocks a significant amount of direct starlight, the degree of polarisation will be enhanced. If the disc is blocking the majority of the star, we will only see light from the visible surface. Based on the observed and theoretical radii we derive the fraction of received flux as $\left(R_{\mathrm{fit}}/R_{\mathrm{evo}}\right)^2\approx1\times10^{-3}$. CS Cha b is far too faint for its temperature, and therefore the low $A_v=0.5$ that we derive is most likely not the line-of-sight extinction. For a star that is occulted by its own disc we expect significant extinction $A_v\sim100$.
The light that we observe could either be direct starlight from a small area or even purely reflected light from the disc, but given the low $A_v$ we are most likely looking at scattered light. Another argument in favour of the highly inclined disc geometry is the enhancement of the OI lines with respect to the H$\alpha$ line. The forbidden OI lines trace outflows that are ejected perpendicular to the disc. A highly inclined disc will also provide less obstruction of the outflow emission.

We determine the mass of CS Cha b with three different methods. In the first we derive the posterior of the mass from the estimated radius of the evolutionary model ($R_{\mathrm{evo}}=1.26\pm0.04$ $R_{\odot}$) and the surface gravity. From this posterior distribution we derived $M=0.25_{-0.18}^{+0.46} M_{\odot}$. The mass can also be derived from the relation between mass and spectral type \citep{baraffe1996mass}. In Appendix \ref{ap:spec_type} we determine the spectral type of CS Cha b, which is estimated to be between M1 and M8. From the spectral type we determine the mass to be between 0.1 $M_{\odot}$ and 0.6 $M_{\odot}$. Although we need to keep in mind that this relation has been derived from main sequence stars and not pre-main sequence objects, such as CS Cha b. Finally the mass can also be estimated from evolutionary tracks \citep{baraffe2015btsettl}. We include all tracks from the ages of 1 Myr to 5 Myr and find all masses for which their $T_\mathrm{eff}$ and $\log{g}$ fall within the 3$\sigma$ contours of the temperature and surface gravity MCMC posterior. This results in a mass range of 0.2 to 0.4 $M_{\odot}$. All different mass estimates place CS Cha b well into the stellar regime. This would make CS Cha B (note the capital letter `B') an M-type star instead of an exoplanet or a brown dwarf.

The derived fundamental parameters allow us to estimate the accretion rate of CS Cha B. For low-mass stars, the accretion rate can be determined either from the line flux \citep{rigliaco2012accretion} or the line width \citep{natta2004accretionwidth}. The relation from \citet{rigliaco2012accretion} implies a mass accretion of $\dot{M}=1.5\times10^{-11\pm0.6} M_{\odot}\mathrm{yr}^{-1}$ after taking all posterior distributions into account, and having corrected for the fraction of received flux. The flux correction does assume a spherical accretion scenario with a 100\,\% filling fraction. If we do not correct for the area we find an accretion rate of $\dot{M}=2.5\times10^{-15\pm0.7} M_{\odot}\mathrm{yr}^{-1}$. 

We can also determine the accretion rate from the empirical relation between the FW10 of the H$\alpha$ line and mass accretion, which is valid for objects with a FW10 of $\ge 200 \mathrm{kms}^{-1}$ \citep{natta2004accretionwidth}. This is the case for CS Cha B, which has a measured FW10 of $298\pm43$ kms$^{-1}$. From the FW10 we determine a mass accretion rate of $\dot{M}=1\times10^{-10\pm0.6} M_{\odot}\mathrm{yr}^{-1}$. This is consistent with the accretion rate that was corrected for the fractional flux, lending additional credibility to the edge-on disc interpretation. The average accretion rate is $\dot{M}=4\times10^{-11\pm0.4} M_{\odot}\mathrm{yr}^{-1}$.

\section{Discussion and conclusion}
Our results show that the previous mass estimate for CS Cha B is too low and that this companion is most likely an accreting M-dwarf with an edge-on disc and an outflow. The edge-on disc explains the strong attenuation of the stellar light and the enhanced line ratios between the forbidden OI lines and H$\alpha$. This also agrees with the interpretation of G18 that the strong polarisation signal is caused by material around the companion. Polarisation in combination with accretion and outflow tracers leave little doubt that CS Cha B is a low-mass star with a disc, accretion, and outflows. 

There are other young objects that have been observed with edge-on discs around them, which were first thought to be substellar companions. A prime example is FW Tau b \citep{bowler2014fwtau}, which is quite similar to CS Cha B. Just like CS Cha B, FW Tau b has a low optical and NIR brightness, and a high OI to H$\alpha$ line emission ratio. And both objects do not have low-temperature photospheric features. \citet{bowler2014fwtau} derive a K-band extinction of 2.5-5.5 by comparing the spectrum of FW Tau b to unobscured M dwarfs. Analogous to the case of CS Cha B, FW Tau b was originally thought to be a planetary mass object ($\sim$ 10\,M$_\mathrm{Jup}$) based on its NIR photometry \citep{Kraus2014FWTau}. It was then later shown using resolved ALMA measurements of the Keplerian motion of disc material around the object that it is in fact a low-mass stellar object \citep{wu2017FWTau, mora2020FWTau}. Another example is TWA 30B which is 5 mag fainter in K-band than TWA 30A, which is of even earlier spectral type \citep{looper2010tripple}. 

Strong spin--spin misalignment between multiple systems in stellar clusters is a predicted outcome of hydrodynamic simulations (see e.g. \citet{Bate2018}). Several multiple systems with resolved circumstellar discs showing strong misalignment have indeed been observed, such as for example the HK Tau system \citep{Stapelfeldt1998HKTau, Jensen2014HKTau} or the PDS 144 system \citep{Perrin2006PDS144, Hornbeck2012PDS144}.
However, CS Cha is now one of the few hierarchical triple systems in which all components show active accretion and outflows in which the inner binary has a circumbinary disc \citep{ginski2018cscha}. Active mass transfer between the components and their discs and stellar winds or outflows can have a profound impact on the dynamical evolution and stability \citep{toonen2016tripple}. Follow-up investigation of this system could lead to further insights into the formation of triplets, which occur relatively often \citep{tokovinin2008tripple,tokovinin2014btripple}, and their dynamical evolution.

Currently, CS Cha B is missing  spectroscopic coverage from the NIR to mid-IR (MIR). Future observations targeting these ranges with multi-wavelength imaging or polarimetry could be used to constrain the disc geometry and its grain population. This makes CS Cha B an interesting target for the spectrographs of the Near InfraRed Spectrograph (NIRSpec) and the Mid-Infrared Instrument (MIRI) onboard the James Webb Space Telescope (JWST). Future observations at higher spectral resolution in the optical could provide more information on the outflow emission lines by better constraining the velocity \citep{simon2016slowwinds}, and could possibly be used to measure a spatial offset \citep{bonnefoy2017spectroastrometry}.

The observations of CS Cha B show that we should be very cautious in our classification of faint companions, as they are not necessarily brown dwarfs or planetary mass companions, and that it is important to use spectroscopy over a broad wavelength range to derive their fundamental parameters.

\begin{acknowledgements}
Support for this work was provided by NASA through the NASA Hubble Fellowship grant \#HST-HF2-51436.001-A awarded by the Space Telescope Science Institute, which is operated by the Association of Universities for Research in Astronomy, Incorporated, under NASA contract NAS5-26555.

A significant part of this work was performed when RGvH was affiliated to ESO. RGvH thanks ESO for the studentship at ESO Santiago during which part of this project was performed.

This work is based on observations collected at the European Organisation for Astronomical Research in the Southern Hemisphere under ESO program 0103.C-0524(A).

JB acknowledges support by Fundação para a Ciência e a Tecnologia (FCT) through the research grants UID/FIS/04434/2019, UIDB/04434/2020, UIDP/04434/2020 and through the Investigador FCT Contract No. IF/01654/2014/CP1215/CT0003.

I.S. acknowledges funding from the European Research Council (ERC) under the European Union's Horizon 2020 research and innovation program under grant agreement No 694513.

\end{acknowledgements}

%
%

\bibliography{references}
\bibliographystyle{aa} %

\begin{appendix}

\section{Observations}
Table \ref{tab:obs} lists the observations that were used to fit the model spectra.
\begin{table}[]
\caption{Observations of CS Cha.}
\begin{tabular}{llcc}
\hline
\hline
Epoch & Instrument & Filter/Mode & Integration time [s] \\ \hline
1998.1339 & WFPC2 & F606W & 108 \\
1998.1339 & WFPC2 & F814W & 108 \\
2006.1311 & NACO & $K_s$ & 1140 \\
2006.1311 & NACO & $L_p$ & 1140 \\
2017.1311 & SPHERE & $J$ & 1794 \\
2017.4617 & SPHERE & $H$ & 1700 \\
2019.2973 & MUSE & NFM & 2056 \\
\hline
\end{tabular}
\label{tab:obs}
\end{table}

\section{Spectral type determination}
\label{ap:spec_type}
In this section we compare the spectrum of CS Cha B to those of template M-type stars to derive a spectral index. The templates are taken from the  Sloan Digital Sky Survey’s Baryon Oscillation Spectroscopic Survey \citep{kesseli2017templates}. This template library provides a wavelength coverage from 3650 to 10200 \AA\, at a resolution of $R\geq2000$, which makes it ideal to compare to the extracted MUSE spectrum.

\begin{figure}[ht]
\includegraphics[width=\columnwidth]{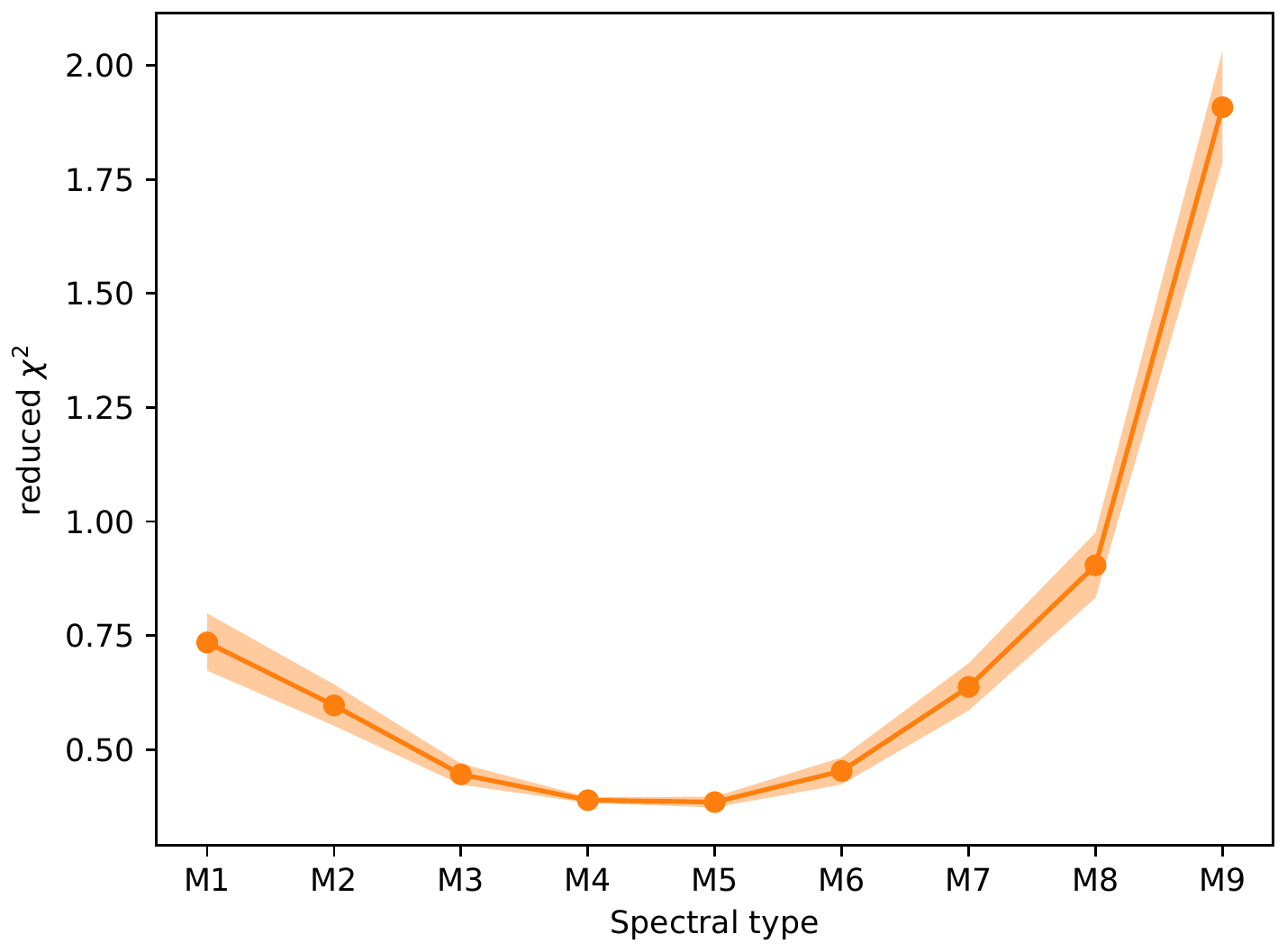}
\caption{ Reduced $\chi^2$ statistic vs. spectral type. The filled region shows the effect of the derived extinction and reddening parameters from the MCMC analysis. The minimum is achieved at a spectral type of M4 or M5. Due to the large error on the measured spectrum we can only exclude the M9 spectral index.}
\label{fig:chi2_templates}
\end{figure}

\begin{figure*}[ht]
\includegraphics[width=\textwidth]{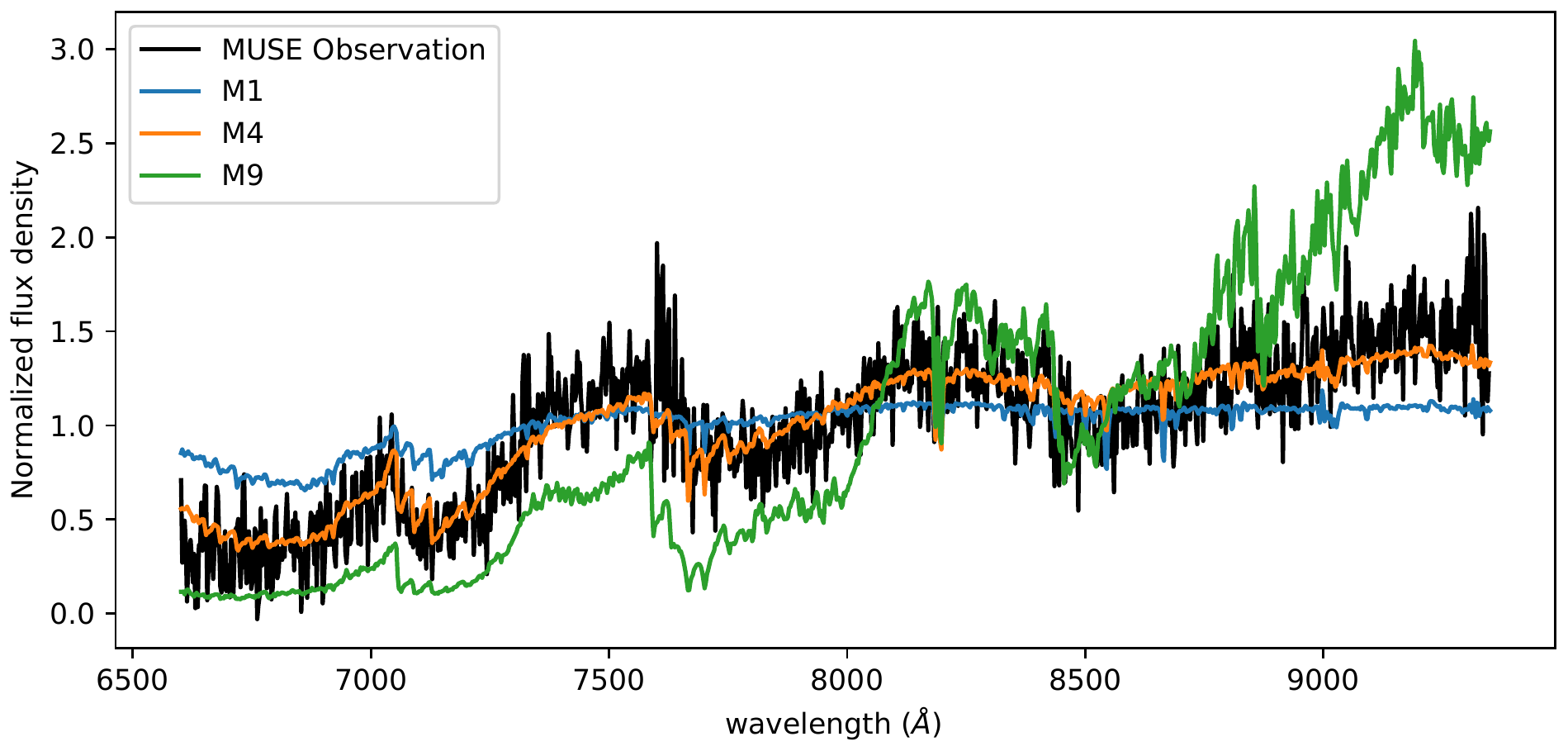}
\caption{ Comparison between the observed spetrum of CS Cha B with MUSE and template M star spectra from \citet{kesseli2017templates}. The best-fitting spectrum is a mid-M-type template, but due to the error on the observed spectrum, only the M9 template can be excluded.}
\label{fig:template_comparison}
\end{figure*}

To find the best-fitting spectral index we calculated the reduced $\chi^2$ statistic,
\begin{equation}
    \chi^2_\nu = \frac{1}{N-\nu}\sum_i \frac{(y_i - d_i)^2}{\sigma_i^2}.
\end{equation}
Here $\chi^2_\nu$ is the $\chi^2$ statistic for $\nu$ degrees of freedom, $N$ the total number of data points, $y_i$ the template, and $d_i$ the data for wavelength $\lambda_i$. Each spectrum is normalised by its integral over wavelength to correct for the different flux scales between the observation and templates before calculating the $\chi^2$ statistic. We also include the effect of the measured extinction and reddening by taking 100 samples from the $A_v$ and $R_v$ posterior and independently calculate the $\chi^2$ for each pair of parameters. In Fig. \ref{fig:chi2_templates} the results of the fit can be seen. The fit indicates that a M4 or M5 template best matches the spectrum. However, there is a significant caveat: the 68\% confidence interval for a $\chi^2$ fit with  1 degree of freedom (spectral index) is $\Delta \chi^2=1.0$. This means that only the M9 template can be excluded; all other templates are statistically indistinguishable within a 68\% confidence interval. The several templates are shown together with the observations of CS Cha B in Fig. \ref{fig:template_comparison}.

\section{The MCMC model}
\label{ap:mcmc}
We used a Bayesian framework to derive the posteriors of the model. At the core of the Bayesian approach are the likelihood calculations which can be used to determine the posteriors through Bayes theorem, 
\begin{equation}
    P(M|D) \propto P(M) P(D|M).
\end{equation}
Here $P(M|D)$ is the posterior of $M$ under the data $D$, $P(M)$ is the model likelihood that is usually called the prior likelihood, and $P(D|M)$ is the data likelihood given a model $M$. We want to determine $P(M|D)$ and simultaneously maximise the probability. This effectively means we will maximise $P(D|M)$. We define our model as
\begin{equation}
    P(D|M)  = \prod_i p_i \prod_j p_j.
\end{equation}
Each $p_i$ is the probability of a measurement, while each $p_j$ is the probability of an upper limit. Each $p_i$ is defined as
\begin{equation}
    p_i \propto \exp{\left[-\frac{1}{2} \left(\frac{x_i - x_{i, \mathrm{model}}}{\sigma_i}\right)^2\right]}.
    \label{eq:prob}
\end{equation}
Here $x_i$ is either a photometric point or a spectral bin, $\sigma_i$ is the measurement error, and $x_{i, \mathrm{model}}$ is the evaluated model. The upper limits are treated in a similar fashion. The likelihood approach requires us to model each observation with a probability density function (PDF). For each upper limit that we include we should therefore also include the PDF that has been used to derive that upper limit. In the case of the two photometric points of G18 the quoted upper limits were zero mean Gaussian distributions with the upper limit as the standard deviation,
\begin{equation}
    p_j \propto \exp{\left[-\frac{1}{2} \left(\frac{x_{j, \mathrm{model}}}{\sigma_j}\right)^2\right]}.
\end{equation}
Here $x_{j, \mathrm{model}}$ is the evaluated model and $\sigma_j$ is the standard deviation.

Maximising Eq. \ref{eq:prob} is the same as finding the maximum of the logarithm of the equation,
\begin{equation}
    \hat{M} = \argmax_{M} P(M|D) = \argmax_{M} \log{P(M|D)}.
\end{equation}
Here $\hat{M}$ is the model with the highest likelihood. The log-likelihood is
\begin{equation}
    \log{P(D|M)}  = \sum_i \log{p_i} + \sum_j \log{p_j},
\end{equation}
and by substituting the individual probabilities we arrive at our final log-likelihood function,
\begin{equation}
    \log{P(D|M)}  = -\sum_i \frac{1}{2} \left(\frac{x_i - x_{i, \mathrm{model}}}{\sigma_i}\right)^2  -\sum_j \frac{1}{2} \left(\frac{x_{j, \mathrm{model}}}{\sigma_j}\right)^2.
\end{equation}

\begin{figure*}[ht]
\centering
\includegraphics[width=1\textwidth]{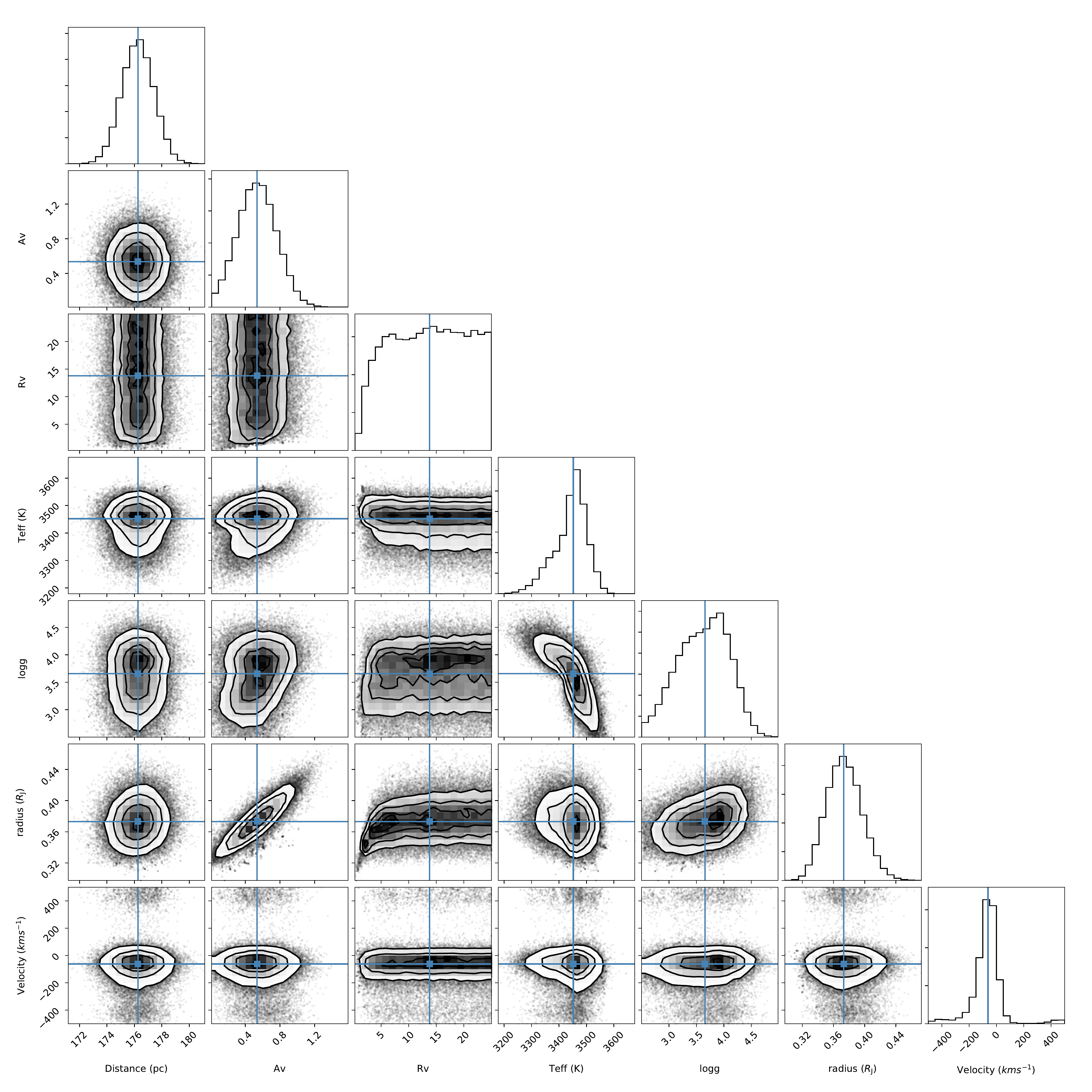}
\caption{ Corner plot of the MCMC sampling. We see a correlation between the surface gravity and the effective temperature. There is a strong correlation between the extinction ($A_v$) and the radius. The $R_v$ is not well constrained and only a lower limit can be determined.}
\label{fig:emcee_corner}
\end{figure*}

\end{appendix}

\end{document}